\documentstyle[prl,aps,multicol,psfig]{revtex}

\begin{document}

\title{Anomalous density dependence of static friction in sand}

\author{Viktor K. Horv\'ath$^1$, Imre M. J\'anosi$^2$\cite{fn0},
and P\'eter J. Vella$^3$}

\address{$^1$ Department of Atomic Physics, E\"otv\"os University,
Puskin utca 5-7., 1088 Budapest, Hungary \\
$^2$ H\"ochstleistungsrechenzentrum, KFA-J\"ulich, 52425 J\"ulich, Germany\\
$^3$ Laboratory for Information Technology, E\"otv\"os University,
M\'uzem k\"or\'ut 6-8, 1088 Budapest, Hungary }

\date{to appear in PRE August 1996}

\maketitle

\begin{abstract}
We measured experimentally the static friction force $F_s$ on the surface
of a glass rod immersed in dry sand.
We observed that $F_s$ is
extremely sensitive to the closeness of packing of grains.
A linear increase of the grain-density
yields to an
exponentially increasing friction force.
We also report on a novel periodicity of $F_s$
during gradual pulling out of the rod.
Our observations demonstrate
the central role of grain bridges and arches in
the macroscopic
properties of granular packings.
\end{abstract}


\begin{multicols}{2}

Recently, there is a considerably increasing interest
in  static and dynamic properties of
granular materials\cite{mon1,mon2,mon3}.
One of the basic problems is to attribute the macroscopic
properties of granular materials to the microscopic
characteristics of grains.
A well known example is a sandpile which exhibits an inclined
free surface of a specific angle, the angle of repose.
Individual grains build up the slope by the action of
microscopic frictional forces at the particle contacts.
It is remarkable that this angle of repose still cannot be
computed from the microscopic characteristics of grains.
Although current efforts on large scale
computer simulations\cite{Gran_comm}
have revealed promising details of
dynamical processes involving grains,
experimental investigations of granular packings still
play pioneering role in the exploration
of the physical nature of granular
materials.

There are a few general laws relating
the magnitude of friction force to principal macroscopically
observable variables\cite{fric1,fric2,fric3}.
The basic one states that the friction force $F$
depends on the normal force $L$ acting between the surfaces
as $F=\mu L^n$, where
$\mu $ is the coefficient of friction, and
$n\approx 1$ is the so called load index.
The force $F_s$ required to start sliding is greater than
the force $F_d$ maintaining the motion.
Assuming that $n$ does not depend on the dynamics,
this difference has led to the notion of two coefficients:
One for the static friction $\mu_s=F_s/L^n$
and one for the dynamic friction $\mu_d=F_d/L^n < \mu_s$.
Another law, usually referred as Amonton's law,
states that the friction force
is independent of the apparent contact area.
Both quantitative laws are generally well obeyed, exceptions
to them are rarities, and the deviations remain
within a few percent in most cases
of sliding coherent {\it solid bodies}\cite{fric1,fric2,fric3}.

One of the first formulations of a macroscopic
friction coefficient for {\it granular
materials} is attributed to Coulomb (1773), who
defined it as the tangent of the angle of repose.
Recent measurements\cite{csonak}
support the idea that horizontal granular layers sliding on
each other obey the general friction laws, too. However,
some reports on related experiments suggest that granular
materials may have very unusual properties, and one should consider
carefully the application of such ``fundamental'' principles,
like friction laws.
For instance,
Allen\cite{all} found that the angle of
repose of different granules depends strongly on the fractional
concentration $C$, which is defined as the total volume of the
grains over the total volume of packing (grains and voids).
Also, the interaction on powder-wall contacts
generally do not obey these laws\cite{brisc}.
In some cases the friction was found to be directly proportional
to the apparent contact area, and quite sensitive to the normal
load\cite{brisc}.
An important feature of the powder-wall contacts is
that the grains have some freedom of movement with respect
to their neighbors, therefore the particles may contribute
more or less independently to the overall friction along the wall.

In order to get insight into this problem,
we performed measurements with a
simple but sensitive experimental setup (Fig.~1).
Over a long period of time similar setups
have been used
occasionally
for related experiments\cite{jen,wieg,dm} (see below), however to
our best knowledge, none of them revealed the phenomena
presented in this work.
\begin{figure}[h]
  \begin{center}
    \leavevmode
     \psfig{figure=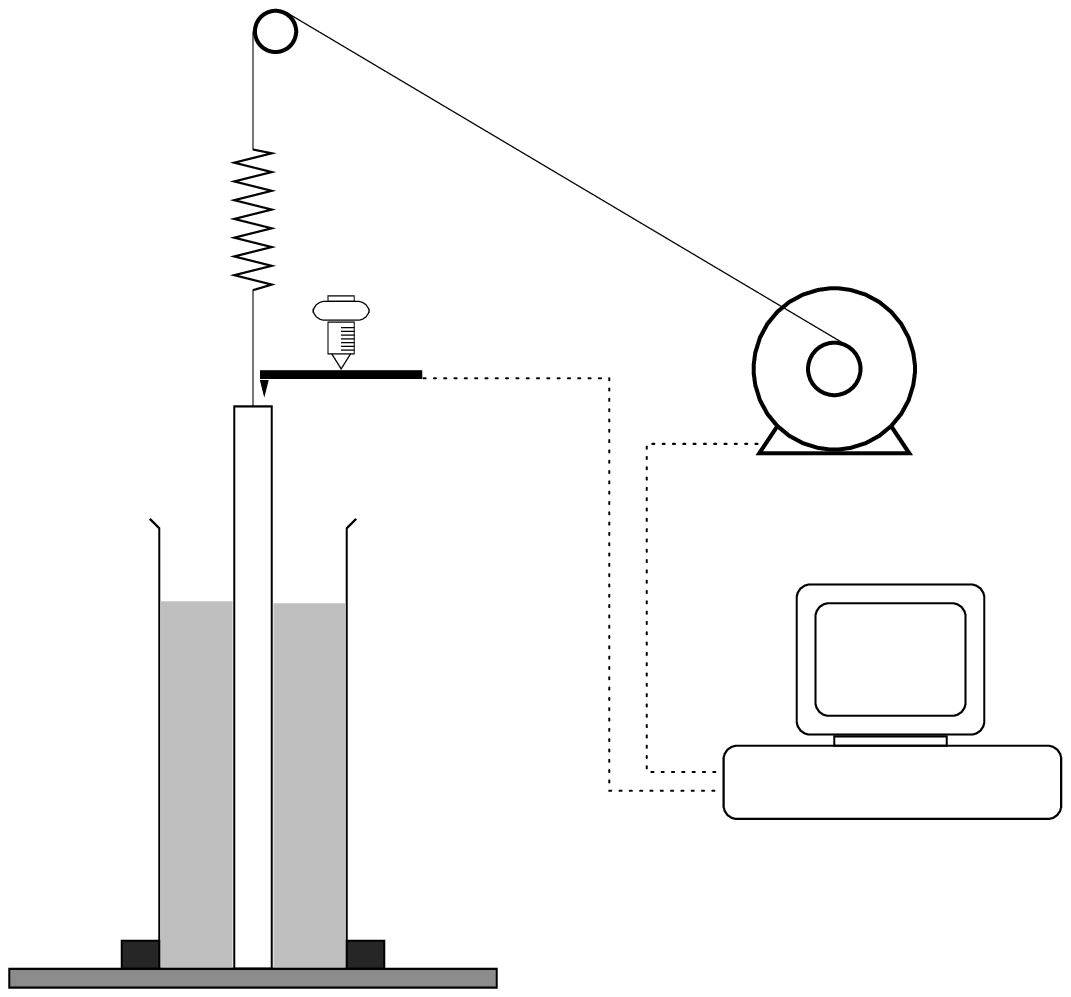,width=7cm}
  \end{center}
{\ FIG. 1. Schematic of the experimental setup. A glass rod is pulled
out from sand by a calibrated steal spring.
The elongation of the spring is adjusted and measured by
a computer controlled stepping motor.
The slip of the rod is detected by closing an electrical
circuit between the top of the rod and an external obstacle.
The gap between the rod and the obstacle is adjustable precisely
by a micrometer screw.
}
\end{figure}

\begin{figure}[hf]
  \begin{center}
    \leavevmode
     \psfig{figure=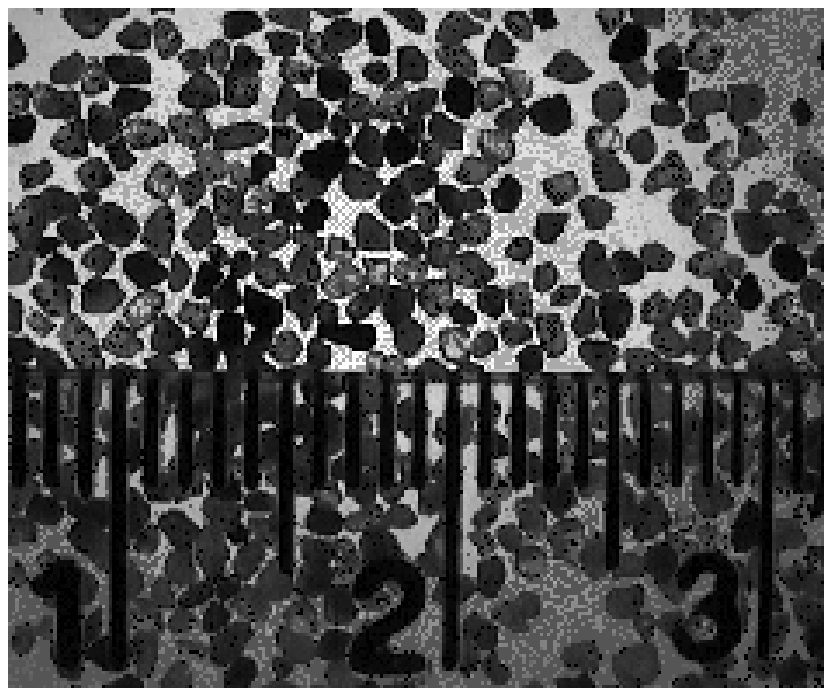,width=7cm}
  \end{center}
{\ FIG. 2. Videomicrograph of the sand grains with a charactersitic size of $\sim 1$ mm
 used in the experiment. The distance between two minor markers on the
ruler is $1$ mm.
The sand with a typical grain-size of $\sim 0.08$ mm (not shown) has
similar irregular shapes.
}
\end{figure}
Our setup consists of a
quartz-glass cylinder with an inner diameter of $D=51.5$ mm,
which is vertically fixed on a precision balance.
A glass rod of diameter $d=10.0$ mm is hung centrally
into the cylinder.
A thin copper sheet for electrical
contact and a small hook
were glued to the top of the rod.
The pulling out force is transmitted
by a calibrated steal spring connected to the hook.
This spring
can be elongated by a
computer controlled stepping motor.
The spring and the axis of
the motor is connected by a special twisted scale-wire of
high flexibility and very low tensile modulus.
An important part of the setup is an obstacle fixed
{\it centrally} above the rod to an outer frame.
The distance between the top of the rod
and the obstacle can be adjusted precisely by a
micrometer screw.
If the upward moving rod hits the tip of the obstacle,
an electrical circuit signals to the computer to stop the motor.
The measured quantity is the number of
steps performed by the stepping motor, from which the elongation
of the spring and then the force can be obtained.

In order  to measure  the
absolute value of  the pull-out
force, we  had   to  define  a   reference  point  for   the
experiments.   Prior  to  each  experiments  we measured the
weight  $W_0$  of  the  cylinder  filled with the same
amount of sand  as used in  the actual experiment.
Next, we
removed the  sand, inserted  the rod  into the  cylinder and
filled the cylinder again with the sand.
Then
we rotated the motor until the weight of the system had
decreased to $W_0$.
At this position the weight of  the
rod  is  balanced  exactly  by  the  spring-force,  thus we
defined  this  as  the  reference  point
for the given packing.
Further   steps   of   the   motor involved
decreasing weight which was directly  related
to the pulling  force of the  elongated spring.
The calibration of the spring was performed
by measuring the weight decrease from $W_0$ versus the
elongation steps of the spring
with the rod temporarily  fixed to
the cylinder.
According to this calibration, all experiments were
performed  in  the  linear  regime  of  the  spring of
modulus $K=21.3173 \pm 0.0004$ N/m.

We   used   two   types   of   sharp  (irregularly shaped)
quartz-sand, one with a characteristic grain-size  of
1 mm (see Fig.~2), and an other of 0.08 mm. 
Both have the same specific weight
$\gamma = 2.59\pm 0.03$  g/cm$^3$.  
The sand was  thoroughly
washed  and  dried  to  remove  intergranular dust and other
contaminations.
To start an experiment, we
measured the height of the
filling, set the system to the reference position and
moved the obstacle away from the top of the  rod
by a given distance $\delta h$.

The measurements  are  performed   in  two
different  ways.   In the first series of
experiments,  the  spring  is  elongated by the
stepping motor until the rod
hits the obstacle.
Then the spring  is released {\it fully},  and
we move the  rod back  to the   bottom of the cylinder.
We rotate the motor to the reference position and
repeat the  same measurement  again.
Note that the  amplitude
of the vertical motion $\delta h$ was much less than the
average grain
size, therefore no grains could move under the rod.  
Typical
results  are  shown  in  Fig.~3.   Based  on  a  series   of
measurements, we
observed that a given fractional
concentration  $C$  is  not
a precise  control parameter  of the {\it first} pull-out
friction  force $F_1$.
Although we always applied  the same filling procedure,
used the same  amount of sand, and
started an experiment only  if we obtained the  same filling
heights within 2\%,  the result
scattered  within  the range of $F_1\approx
0.9\pm0.5$ N
in the case of Fig.~3.
The mean value and the scatter are based on 10 realizations,
in each case the pulling-pushing cycle was repeated more than 20 times.
\begin{figure}[ht]
  \begin{center}
    \leavevmode
     \psfig{figure=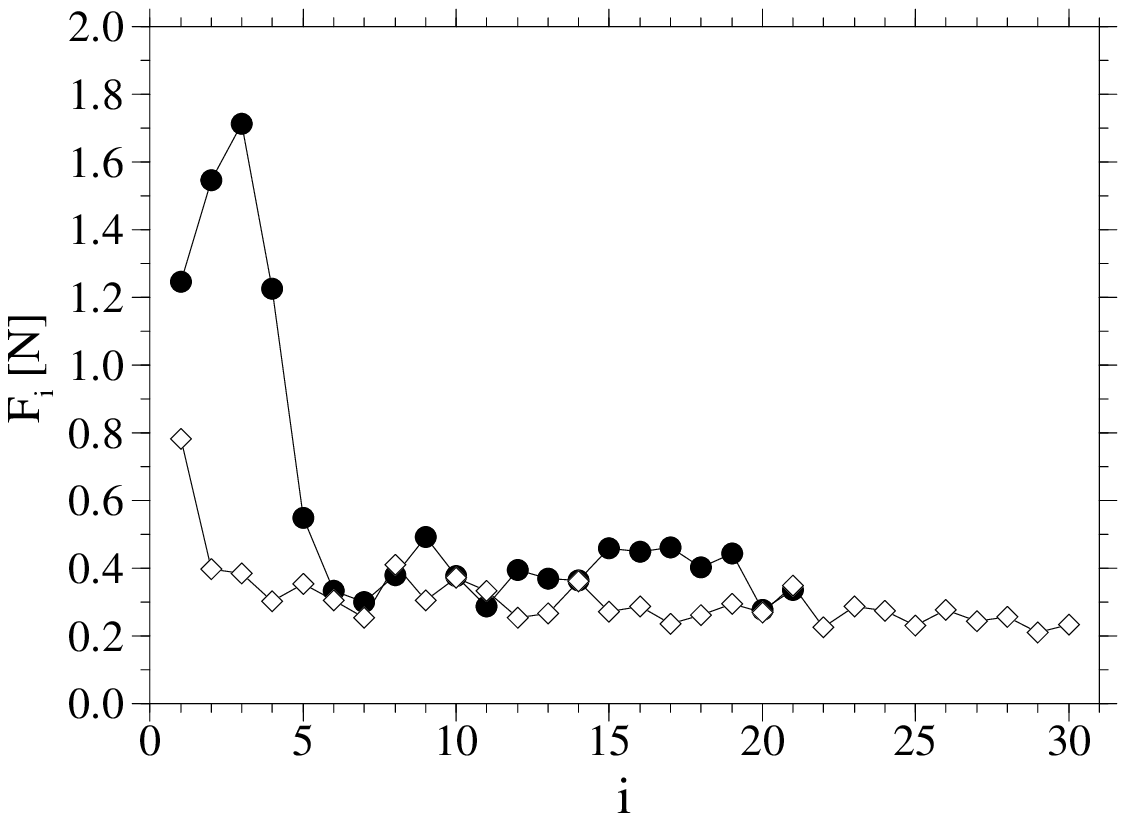,width=7cm}
  \end{center}
{\ FIG. 3. Two series of pulling-pushing measurements with the same amount
of loose packed sand of $\sim 1$ mm grain-size, $m=745$ g.
The vertical axis shows
the static friction force $F_i$ at the $i$th pull, after each
attempts the measuring spring was released and the rod pushed
back to the original position.
The  allowed
 amplitude of the rod motion was
$\delta h = 150\pm 5~\mu$m.
The height of both fillings was $24.4\pm 0.2$ cm
(i.e. the fractional concentration $C=0.590\pm 0.009$).
}
\end{figure}
Also, the  way of  reaching an
asymptotic force can be very different.
Usually we
observed initially a decrease of the friction force $F_i$
at repeated cycles.
 This
``weakening'' is shown by
white diamonds in Fig.~3.  However,
in some cases we observed an opposite tendency during the first
few pull out steps.  Such  a ``strengthening'' is  demonstrated
in Fig.~3 by the first  three filled circles.
We found that  the asymptotic value $F_a$ of
 the pullout force for a given filling-mass
is  independent
from the first pull-out force $F_1$ and depends
only on the  fractional
concentration $C$.
The numerical value of the asymptotic force in Fig.~3
is $F_a=0.28\pm 0.09$ N.

During  our  experiments, we never observed
macroscopic rearrangement  of the  filling, or  some visible
change on the surface.  These would suggest the expansion of
the packing, i.e. macroscopic dilatancy.  Therefore we think
that  the  change  of  the  force  in repeated attempts is a
consequence of microscopic rearrangements along the wall  of
the rod.
Such rearrangements can happen if some small voids  cave
in, or some grains rotate into another position, which  does
not effect the stability of the overall filling.
After several ``polishing'' cycles, probably the structure finds
a static configuration at which
the pullout force shows a saturation.

As for the filling height $(h)$ dependence of the first pullout force $F_1$,
Dahmane and Molodtsof\cite{dm} performed
related experiments at a constant fractional concentration $C$.
They obtained the empirical formula\cite{dm}
$$ F_1(h)=\mu _s P\gamma {ah^3\over (b+h)^2}\quad , \eqno(1)$$
where $\mu _s$ is a rod-coating dependent coefficient
of static friction, $P$ is the perimeter of the rod, $\gamma $ is the
specific weight of the granular material, $a$ and $b$ are
(positive) empirical constants with dimensions of length.
Their main result is that the pullout force $F_1$ does
not depend on the radial position and the shape of the cross section
of the rod\cite{dm}.
We carried out a few independent tests in loose packed sand to check the
filling height dependence of $F_a$ in the range of $10 <z<25$ cm.
Our results are consistent with Eq.~(1),
which involves approximately a linear dependence on $z$, apart from
very shallow fillings.

Next, we performed repeated pullout measurements in a
different way.
The beginning of the experiment was identical to the
procedure described above.
After the first measurement, the spring
was set to the {\it new} reference position
according to the increased height of the rod,
but instead of moving the rod back to the
initial depth, the obstacle
was moved away from the top of the rod again by the
same distance $\delta h$.
Thus the rod was allowed to move out a larger distance
step by step.
Additionally to the previously described weakening and strengthening,
we observed a novel oscillatory behavior.
In Fig.~4, a representative result is plotted for sand
of typical grain size of 1 mm.
Note that halving the distance $\delta h$ between subsequent steps
does {\it not} affect the period length,
but doubles the number of data points in one period.
For relatively loose fillings ($C\approx 0.6$) this period length
is roughly 1/3 of the grain size,
and seemingly does not depend on the filling height in the range
of $9<h<16$ cm.
We could not observe clear periodicity using
the sand of smaller grain size.
Larger packing densities involve much larger
pull-out forces (see below), which
cover this oscillation and make its observation much harder.
It is interesting to note that although the period length
is reproducible for the same conditions,
the initial phase changes randomly from sample to sample.
In Figure 4, we horizontally shifted the second series (empty
diamonds) by
$\Delta h =0.15$ mm
to make the periodicity more transparent.
\begin{figure}[t]
  \begin{center}
    \leavevmode
     \psfig{figure=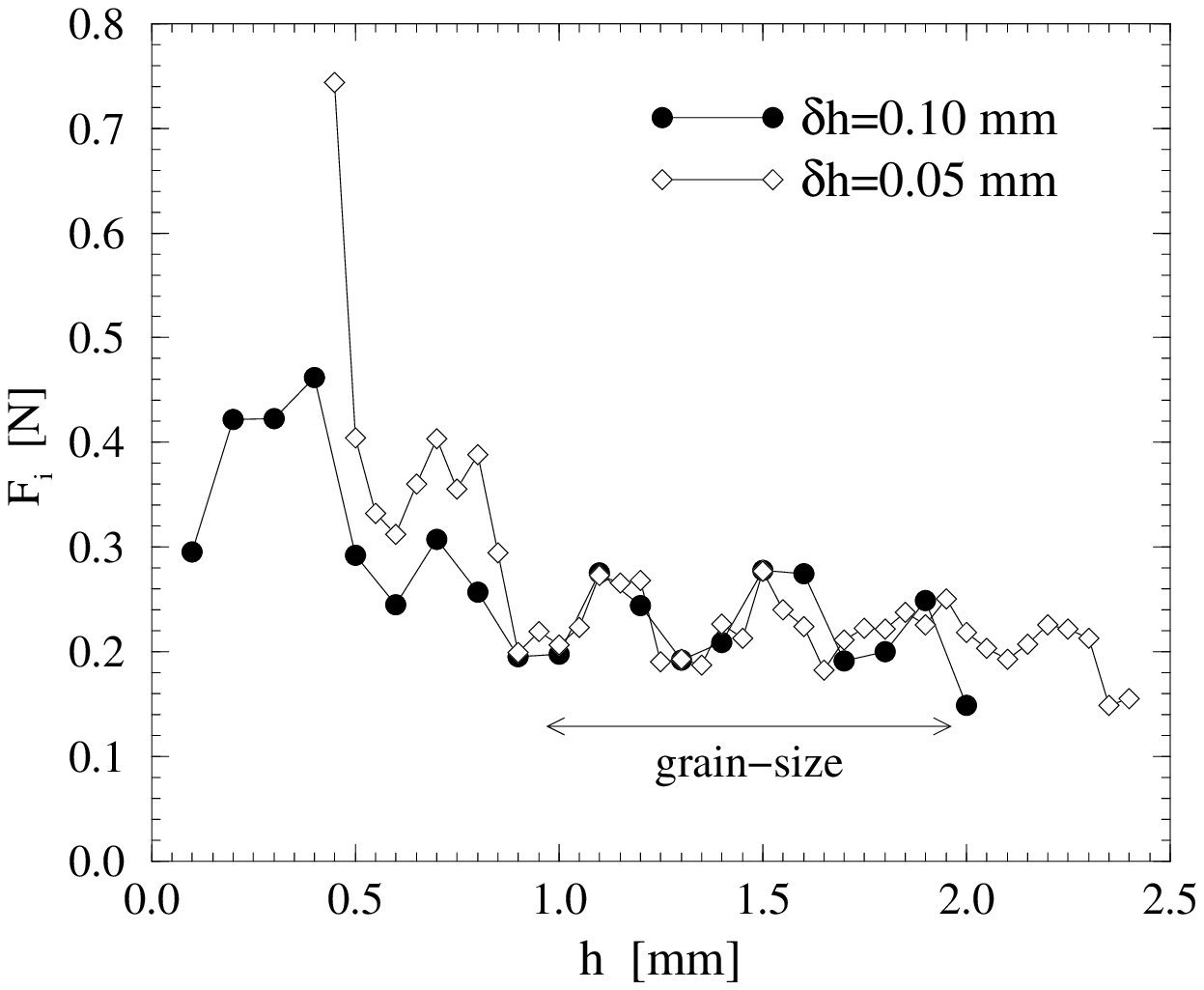,width=7cm}
  \end{center}
{\ FIG. 4. Static friction versus rod height
in a gradual pull-out experiment at two different step sizes.
After each pulling the measuring spring was released, and
a new measurement was performed without changing the rod position.
The filling height of sand of $\sim 1$ mm grain-size was $16.3\pm 0.2~$cm,
loose packing case
($C\approx 0.578$). The series of
step size $50\pm 1~\mu$m
(empty diamonds) is shifted horizontally to indicate the
constant period length. The arrow shows the typical grain-size.
}
\end{figure}
This random phase-shift can be related to the
opposite tendecy of the observed weakening and strengtheing
 in the previous experiments.
In Figure~4., appart from the oscillation, we can
also
observe a global decrease
of the friction force versus the {\it total distance} of the repeated pull-out
steps.
We can identify two different regimes
depending on the total pull-out distance.
First, within the size of one grain diameter
we observe a fast, non-linear
decay of the friction force, this is in analogy with the weakening
found by the pulling-pushing experiments.
For longer distance of pull-out, the average value of the force
(averaged over one oscillatory period)
decreases continuously further, but much slower than at the
beginning.
On one hand, this slow decay might be related to the decreasing total contact
area.
On the other hand, the appearing free void under the tip of the rod might
contribute
to the decay, giving a freedom grains to move slightly in underneath the rod.
We return to this point below.
We mention here that large fluctuations and sometimes oscillations
were reported also by Meftah et al.\cite{mef}
in a different {\it shearing} experiment of
two-dimensional packings.
Although these
oscillations are apparently
inherent properties of
granular systems, a satisfactory theoretical explanation
of them is missing.

Next we turn to the effect of the closeness
of granular packings.
Already in 1931, Jenkin\cite{jen} reported on
a series of unsuccessful measurements for friction coefficients
of sand. Without providing data, he noted that the {\it closeness
of packing} of the grains was an essential factor in determining
the behavior and the irreproducibility of the results.
Experimentally, we changed the density
of the sand by vibrating
the cylinder vertically with 50 Hz by an
electromagnet. The amplitude of vibration was well below
the fluidization limit.
During compaction, the rod
was kept fixed at the central position.
We followed the increase of density
by detecting the decrease of the height.
Longer time of vibrating resulted in a larger density.
The typical effect of increased
closeness of packing is illustrated in Fig.~5.
During repeated attempts, the pull-out force showed a
weakening tendency in the densified packing again, but its value
saturated at a higher level
than the asymptotic value for the previously measured loose packing.
We stress that strengthening and clear oscillation were never observed
after compacting, which suggests that
strengthening in loose packings might be closely related to the
observed force-oscillations at low densities.
\begin{figure}[hf]
  \begin{center}
    \leavevmode
     \psfig{figure=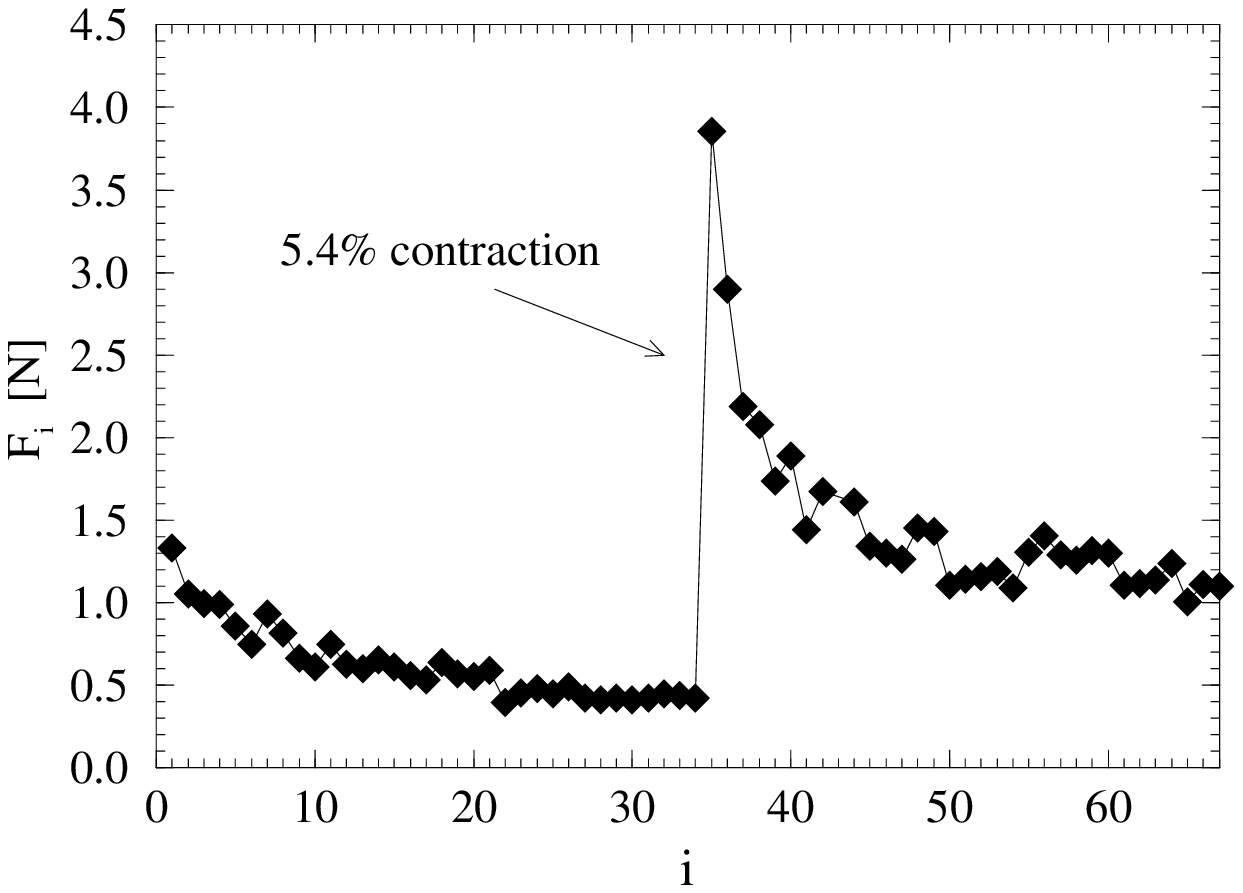,width=7cm}
  \end{center}
{\ FIG. 5. The effect of increased density on the static
friction $F_i$ (pulling-pushing experiment).
At the beginning, the density of the filling was
$\varrho = 1.524 \pm 0.015$ g/cm$^3$
($C=0.588\pm 0.009$), after the
saturation of the force the filling was
compacted to a value of $\varrho = 1.606 \pm 0.015$
g/cm$^3$
($C=0.620\pm 0.009$). The other parameters are the same
than in Fig.~3.
}
\end{figure}

\begin{figure}[hf]
  \begin{center}
    \leavevmode
     \psfig{figure=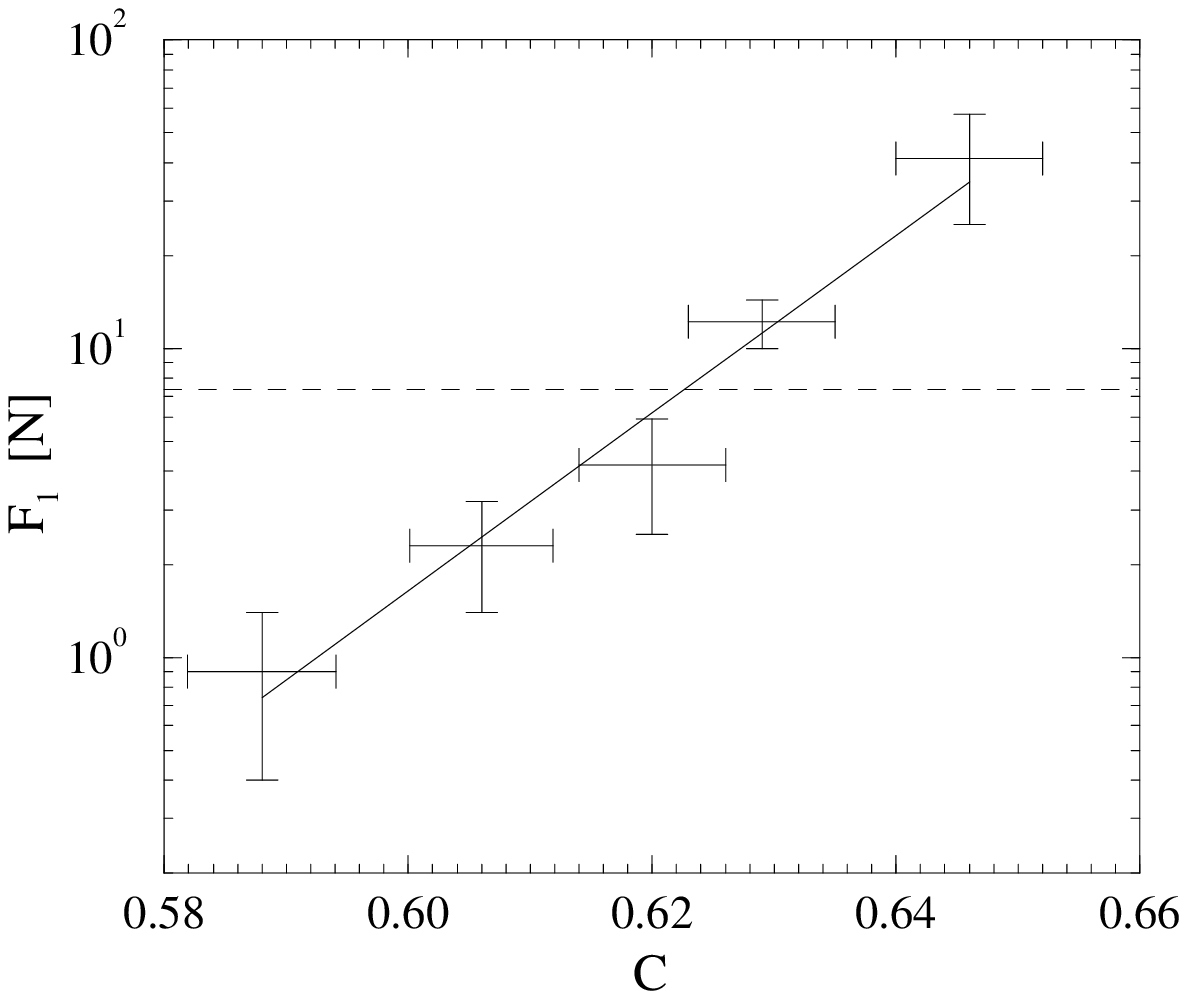,width=7cm}
  \end{center}
{\ FIG. 6. First pull-out force $F_1$ as a function of fractional
concentration $C$. The mass of the
fillings was $745\pm 0.05$ g.
$C$ was obtained by height measurement.
The error bars represent sample to sample fluctuations
of 10 realizations with  $\sim 1$ mm grain-size and
3-5 realizations with the finer sand at each $C$ values. The solid line
shows an exponential fit [see Eq.~(2)], the dashed line indicates
the weight of the filling.
}
\end{figure}
We measured the density dependence of the
first pull-out force using a fixed amount of sand,
the result is shown in Fig.~6.
Surprisingly, the force does  not depend on the  grain size,
both type of sands gave practically the same result.
Apparently the density
dependence is exponential:
$$ F_1(C) =\kappa e^{C / C^* } \quad , \eqno(2) $$
where $C^* \approx 0.015$ and $\kappa $ are
 fitting parameters [$\ln(\kappa )=-39.265$].
Obviously the domain of
validity for Eq.~(2) is bounded by a minimal
and a maximal possible packing concentration, which is
approximately the interval $0.54<C<0.68$ for natural quartz
sands in air\cite{all}.

There is no easy way to attribute the first pull-out force $F_1$
to macroscopic variables.
For small densities of the packing,
the total load on the wall $L$ should be proportional to the
weight of the sand (indicated by a dashed line in Fig.~6),
because in this case
mostly gravity induces normal forces in the system.
Then one may explain the increase of $F_1$
by the fact that densification increases
the real contact area along the surface of the rod, although the
apparent contact area decreases.
This effect is limited by an incerease of average coordination
number in the packing, therefore it can not account for such a
large increase of the friction force.
Moreover gravity itself can not induce larger
normal force than the total weight of the sand.

It is well-known, however, that shearing can induce normal forces
in granular packings, too\cite{tg}.
Indeed, the observation that the friction force can exceed the total
 weight of the
sand (see Fig.~6.) indicates that shear-induced normal
forces dominate for dense packings.

\begin{figure}[hf]
  \begin{center}
    \leavevmode
     \psfig{figure=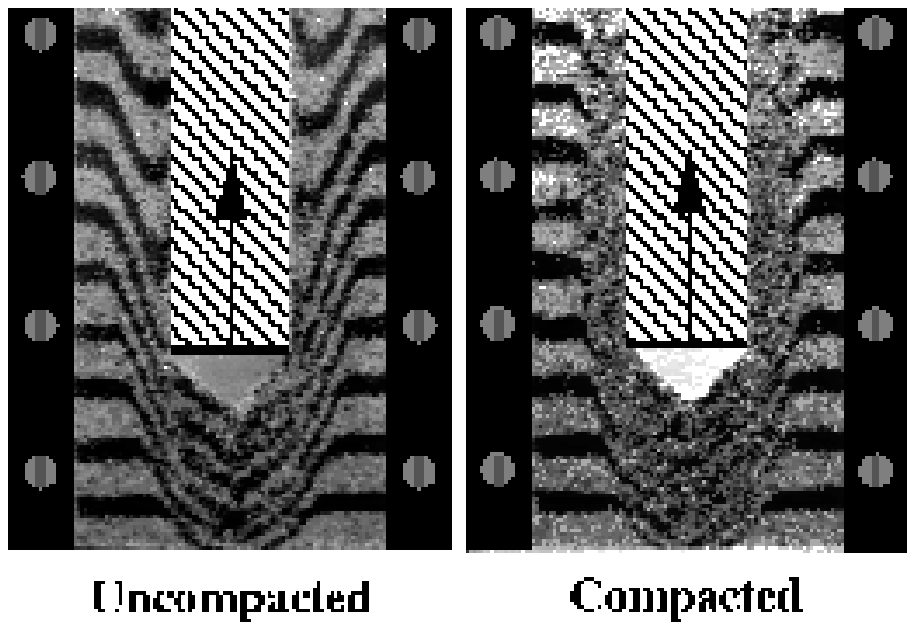,clip=true,width=8cm}
  \end{center}
{\ FIG. 7. Quasi-2D visualization of the local rearrangement of sand due to the
upward movement of the rod.
The black layers formed by the same
type of sand colored with ink.
The internal width of the cell is $120$ mm, the depth is 12 mm.
In both cases we filled the cell to the same height with sand of
average grain size $1$ mm, the compaction was 4.9\% (right picture).
}
\end{figure}
The force distribution
is not homogeneous in granular materials\cite{sci}.
Interparticle forces are
transmitted by a discrete number of
irregular contact paths, the grain bridges or arches.
The dominant feature leading to force chains
is the strong disorder of the packing, which causes
a highly irregular distribution of weights on grains.
Recent observations\cite{sci} show that the force
distribution spans over a wide range, and has an
exponential tail. This indicates that very strong
contact-forces may be present.
At deformations, individual
grain bridges collapse and
could be replaced by new ones.
The bridges can fail in many ways: By the fracture
of a grain or the bounding surface, by slip between
grains or between a grain and the bounding surface.
The slipping mechanism of bridge failure is basically
based on spatial rearrangements. This rearrangement
is connected to the well known phenomenon of
dilatancy,
i.e.~grains need an extra volume to roll or slip
over each other.

In order to visualize 
the movement of sand grains 
while the rod is pulled out,
we built a quasi-2D version of the
experimental setup bounded by two parallel plexiglass plates. Although
the different 
geometry is expected to effect the behavior of the sytem,
we do beleieve that some global aspects of the dynamics
is preserved. To make the local rearrangement of the sand grains more
transparent, we layered colored and normal sand in horizontal strips.
In Fig. 7. we show the resulting pictures after pulling out
the rod by a distance of $h=43$ mm.
We observed that at dense packings
the flow-regime extends to a smaller distance both
longitudinally and laterally, while the mixing inside is much
stronger. (This mixing is not a consequence of compaction, note
the clear separation of layers in the unperturbed regions.) 
This shows that shearing in a compacted assembly results in strong
local rearrengements along the wall of the rod, while moderate
but extended structural changes are characteristical in loose
packings. Since the free void space is reduced in a dense medium,
we can conclude that local rearrengements should be associated
with larger contact forces at higher fractional concentrations.
     
This observation also suggests an explanation for the slow decay
of the average friction force at gradual pullout (see Fig.~4).
If the pullout distance of the rod is larger than the average grain-size,
the void under the tip gives a free volume for grains to move in.
Although the flow under the rod involves only a few grains
at a height of $1-2$ mm, the release of strong local contact
forces at around the tip can result in a macroscopically observable
decrease of the pullout force.

We suggest that local dilatancy furnishes the key
to understand the force strengthening
(Fig.~3), as well as periodicity (Fig.~4) in loose packings.
There is enough free volume inside the packing given
by the interparticle voids, thus bridges
can collapse and build up without
a macroscopic volume expansion. In close packings,
however, similar
local rearrangements are hindered by geometrical constraints,
therefore some grain bridges can support very large forces.
Since we did not observe macroscopic dilatancy in our
measurements, we can conclude that the main bridge
failure mechanism in this case is probably slip at the boundary
surface.

We are grateful to
G. Batrouni, J. Hajdu, H. Herrmann, J. Kert\'esz, F. Radjai,
Gy. Radnai, I. Szab\'o and D. Wolf for
many discussions.
This work was partially supported by the Hungarian
National Science Foundation (OTKA) under grants No.~F17310
and No.~F014967. One of us (V.K.H.) thanks for the
Foundation for Research and Higher Education
for the Zolt\'an Magyary scholarship.

\vfill
\end{multicols} 
\end{document}